\documentclass[aps, prd, twocolumn, showpacs, superscriptaddress, groupedaddress]{revtex4-1}

\usepackage{graphicx}
\usepackage{yfonts}
\usepackage{float}
\usepackage{amsmath}
\usepackage{amssymb}
\usepackage{amsfonts}
\usepackage{soul}
\usepackage{amsthm}
\usepackage{mathtools}
\usepackage{dcolumn}
\usepackage{listings}
\usepackage{subfigure, rotating, bm, array}
\usepackage[utf8]{inputenc}
\usepackage[english]{babel}
\usepackage[pagebackref=false, colorlinks=true]{hyperref}
\hypersetup{
linkcolor=blue,     
citecolor=blue,     
urlcolor=blue}      

\usepackage{float}
\begin{document}
\title{Non-minimal coupling of scalar fields in the dark sector and generalization of the top-hat collapse}
\author{Priyanka Saha$^\dagger$, Dipanjan Dey$^\ddagger$, Kaushik Bhattacharya$^{\$}$}
\email{priyankas21@iitk.ac.in, deydipanjan7@gmail.com, kaushikb@iitk.ac.in}
\affiliation{$^\dagger, ^{\$}$ Department of Physics, Indian Institute of Technology, Kanpur,
Kanpur-208016, India\\
$^\ddagger$Beijing Institute of Mathematical Sciences and Applications, Beijing 101408, China}

\begin{abstract}
In this article we propose a new way to handle interactions between two scalar fields, in the cosmological backdrop, where one scalar field oscillates rapidly in the cosmological time scale while the other one evolves without showing any periodic behavior. We have interpreted the rapidly oscillating scalar field as the dark matter candidate while the other scalar field is the canonical quintessence field or the non-canonical phantom field. A model of a generalized top-hat-like collapse is developed where the dark sector is composed of the aforementioned scalar fields. We show how the non-minimal coupling in the dark sector affects the gravitational collapse of a slightly overdense spherical patch of the universe. The results show that one can have both unclustered and clustered dark energy in such collapses, the result depends upon the magnitude of the non-minimal coupling strength.   

\end{abstract}
\maketitle



\section{Introduction}

The missing mass on the scales of galaxies and galaxy clusters remains an important underlying problem in cosmology. The most common explanation for this missing mass comes from the dark matter sector. Dark matter primarily interacts gravitationally, while other interactions of dark matter, if present, are incapable of making the dark matter constituents visible. Till now we have not experimentally observed any clear signature of dark matter.
Despite considerable efforts, dark matter particles have yet to be directly detected. Over the last few decades, many advanced theories have been proposed to explain dark matter, and how this kind of matter is involved in structure formation. According to the cosmological principle, the universe is homogeneous and isotropic, therefore the formation of large structures from small perturbations needs an explicit theory. While the $\Lambda$CDM (CDM for cold dark matter and $\Lambda$ for the cosmological constant) model is widely accepted, with its greatest success in explaining cosmic microwave background radiation in great detail, it faces many challenges. One of the difficulties of the $\Lambda$CDM model is related to the value of the cosmological constant. The value of the cosmological constant required to fit cosmological data turns out to be much smaller than the value of the constant predicted from particle physics models. The other problem of the standard  $\Lambda$CDM model is related to the cosmological era where $\Lambda$ becomes effective in determining the cosmological dynamics. There is no satisfactory answer to the question of why the energy density ratio due to the cosmological constant and due to dark matter is of order unity, presently. All these problems compelled many workers to propose a dynamic dark energy concept where the late acceleration of the universe is due to the dynamics of some scalar field. 

Various models have been proposed in the literature to address the problems of the $\Lambda$CDM model. Several authors have worked on warm dark matter \cite{Avila-Reese:2000nqd, Colin:2000dn}, self-interacting dark matter \cite{Loeb:2010gj}, the braneworld model \cite{Garcia-Aspeitia:2011yyt}, and also models that modify the gravitational force without introducing the dark matter sector \cite{Sharma:2022fiw}. One of the popular models is the scalar field dark matter (SFDM) model, which presents the dark matter as a scalar field with some potential $V(\phi)$. This approach offers a more fundamental perspective to the problem. There exist a large number of potentials that can be used to design scalar fields as dark matter, \cite{Matos:2008ag, Urena-Lopez:2015gur, Turner:1983he, HWANG1997241, Escobal:2020rfn, Lee:2008ab, Magana:2012xe, Suarez:2011yf, Cedeno:2017sou, Matos:2000ng, Alcubierre:2001ea, Matos:2000ss, Urena-Lopez:2019kud}. The choice of potentials may include the trigonometric or hyperbolic functions. Perhaps one of the simplest forms of the scalar field potential is the quadratic potential, $V(\phi)=\frac{1}{2} m^{2} \phi^{2}$, where $m$ is the mass of the scalar field. SFDM models can also exhibit early structure formation \cite{Suarez:2011yf}. In this paper, we will work with SFDM with the quadratic potential as in that case, the scalar field can produce dust-like behavior when it oscillates around the minimum.

It has been shown that oscillating scalar fields can produce CDM-like cosmological perturbations in the late universe \cite{Brandenberger:1984jq, HWANG1997241}. In \cite{Turner:1983he}, the authors showed that an oscillating scalar field with a quadratic potential can produce features of standard CDM in the late universe. In the presence of a scalar field with a quadratic potential, two important time scales appear. One of these scales is $1/H$, where $H$ is the Hubble parameter. This is the time scale in which cosmological changes become apparent, while the other time scale is related to the oscillation of the scalar field in the quadratic potential,  given by $1/m$. Here $m$ is the mass of the scalar field. Turner showed that (in Ref.~\cite{Turner:1983he}) the scalar field can oscillate many times in a typical cosmological cosmological period if $m \gg H$. Consequently, one can average out the fluctuations of the oscillating scalar field to find out the equations of cosmological dynamics. It was then shown that this averaging process transforms the scalar field sector in such a way that it behaves as a pressure-less fluid, like standard CDM. The period of this field oscillations, $m^{-1}$, is much shorter than the Hubble time, $H^{-1}$. Later on, Ratra gave a much more elegant description of the above process in Ref.~\cite{Ratra:1990me}. 

The mass of the scalar field in the various SFDM models (which may or may not use coherent oscillations of the field) can vary. One of these models is the FDM (Fuzzy dark matter) model where the mass of the scalar field is around \( 10^{-22} \) eV,  which has distinct properties compared to the $\Lambda$CDM model. In FDM models, we have delayed galaxy formation relative to $\Lambda$CDM model predictions while the other predictions remain consistent with cosmological observations for the high redshift galaxies \cite{Hui:2016ltb}. 

Axion-like particles with masses around \( 10^{-21} \) eV can act as  SFDM and address the unsolved problems of dwarf galaxies. It is known there exist difficulties in explaining all the properties of dwarf galaxies within the standard $\Lambda$CDM model paradigm \cite{Marsh:2013}. In general the scalar field dark matter models have a wider mass range of the scalar field, masses can range from \( 10^{-33} \) eV \cite{Escobal:2020rfn, Jesus:2015jfa} to \( 10^{-18} \) eV \cite{Marsh:2015xka}. The lower bound corresponds to the order of the Hubble parameter in the present epoch (\( 10^{-33} \) eV).  The scalar field can also form a Bose-Einstein condensate, explaining the problems related to structure formation. It has been shown that the galactic halos can be described as a spherical Bose-Einstein condensate of scalar particles characterized by a coherent scalar field. 

In this article, we will first like to present the spherical collapse model which involves the coherently oscillating scalar field acting as CDM. Our main purpose in this paper will be related to the investigation of spherical collapse in cosmological models where the oscillating SFDM candidate has a non-minimal coupling with the dark energy constituent. Non-minimal coupling in the dark sector is an attractive idea and lately, there are some relevant works related to this topic as can be seen from Refs. \cite{Chatterjee:2021ijw, Hussain:2022osn, Bhattacharya:2022wzu, Hussain:2022dhp, Hussain:2023kwk, Debnath:2024urb}.
In the present case, the dark energy constituent will be the quintessence field or the phantom field. In the spherical collapse model, an overdense spherical patch initially starts expanding at a different rate from the background universe due to its difference in density. Gravity prevents the over-dense region from expanding forever and consequently the overdense region starts to collapse after some time. From pure general relativistic considerations, this collapse ultimately proceeds towards a spacetime singularity. On the other hand, from phenomenological intuition, we know these kinds of collapses ultimately virialize to produce galaxy clusters or individual galaxies. 

Previously there have been some attempts \cite{Saha:2023zos, Saha:2024xbg} to tackle the general relativistic problem of spherical collapse in a two-component system, where one component is made up of a pressure-less fluid, specifying the dark matter sector, and the other component is made up of a quintessence like scalar field, specifying the dark energy sector. In these models, the general relativistic treatment is valid up to the time of virialization, when the collapse stops. General relativistic treatment of spherical collapse with two components, where the pressure inside the collapsing spherical surface does not vanish, requires two kinds of spacetimes which are matched at some radial distance on a time-like hypersurface. In general, the internal spacetime is a closed Friedmann-Lemaitre-Robertson-Walker (FLRW) spacetime and the outer one is the generalized Vaidya spacetime. In \cite{Saha:2024xbg}, the fluid representing dark matter and the scalar field, acting as a dark energy constituent, were non-minimally coupled. In such a model of spherical collapse, one can have clustered dark energy. 

There already exists a collapse model with an oscillating scalar field \cite{Guzman:2003kt}. Traditionally, dark energy is not included in the spherical collapse model, because of the homogeneity of dark energy which prevents it from clustering during collapse. Obtaining the influence of dark energy in structure formation is  an important task as dark energy density is the dominant component of energy density in the present universe. It is expected to have some impact on the formation of dark matter structures in cosmology. 

In our work, we have considered a two-component system described by two scalar fields: one scalar field constitutes the dark matter sector and another scalar field specifies the dark energy sector. An oscillating scalar field will model the dark matter as we have already discussed, while dark energy will be modeled by quintessence or phantom-like scalar fields. First, we will show how an oscillating scalar field can act as a pressure-less fluid so that it can behave as dark matter. Using this scalar field as dark matter we will then work out the spherical collapse model with interacting dark energy to get the collapse dynamics of the system. The background expansion of the universe will be based on the dynamics of the spatially flat FLRW metric. The dynamics inside the overdense region will be based on the closed FLRW metric. In our case, the closed FLRW metric will be matched with a generalized Vaidya spacetime. It will be assumed that the collapse occurs at a length scale much smaller than the cosmological length scale (where dark energy dominates over dark matter, driving accelerated expansion), but is significantly larger than the astrophysical galactic scale. Therefore, we focus on an intermediate galactic cluster scale to examine the influence of dark energy on the structure formation problem. A distant observer, observing the collapse, will only observe the external Vaidya spacetime. Both the background cosmological sector and the inner region of the overdense patch will contain scalar field dark matter with non-minimally coupled dark energy. 

Previously authors have calculated and determined the mass of SFDM to be around [\( 10^{-21}-10^{-23} \)] eV \cite{Marsh:2013,  Matos:2000ss}. Considering this mass range, more work in the SFDM sector has also been done as can be seen from \cite{Guzman:2003kt, Suarez:2011yf, Hu:2000ke}. In the present work, we have not specifically considered the mass \( m \) of the oscillating dark matter constituent to be in this range. Instead, we will maintain a small constant value of \( m \gg H_i \), where $H_i$ is the initial value of the Hubble parameter of the collapsing spacetime, and examine how the mass value affects the conditions for collapse. We have worked with a model where \( m/H_i \) is kept around \( 10^4 \). In our work, we have noticed that the value of the mass of the dark matter constituent significantly affects the regions of the initial condition which gives rise to gravitational collapse.
Outside the specified region of initial values, the initial over-dense region will not collapse and it will expand forever. It is also observed that the initial conditions also gets affected by the type of scalar field (quintessence or phantom) used for modeling dark energy. 

Since we are considering an oscillating field to model the dark matter, we will work with the time-averaged values of the components.
This time averaging works efficiently when the mass of the dark matter constituent is assumed to be much much greater than the expansion rate $H$ of the overdense spherical patch as shown in Ref.~\cite{Ratra:1990me}. In this paper, we have used the units \( 8\pi G = c = 1 \), where $G$ is Newton's universal gravitational constant and $c$ is the speed of light.  In this units both $m$ and $H$ are measured in ${\rm cm}^{-1}$. Unlike the results in natural units, our units do not give mass (or energy) units to both $m$ and $H$. To compare our results with the standard ones we will convert our results to natural units. 

Before we end the introduction it will be useful to remind the reader about the usage of the phrase {\it mass of the scalar field}. Many authors use this phrase, what they mean is the mass of the quantum of the scalar field. That quantum of the scalar field is a particle and one can associate a concept like mass to it. This association of mass of the quantum of the scalar field comes out straightforwardly while using natural units. In the Geometrized units, one only has a length scale attached to $m$ in the potential $m^2\phi^2/2$. This length can be associated with an energy scale when we have a finite $\hbar$ and $c$. This association can be understood from the relationship of the Compton wavelength with the electron mass. In our present work, we do not quantize the theory, and we work with conventional Geometrized units. Although we do not quantize the scalar field sector, we will still call $m$ to be the mass of the dark matter particle which has a unit of ${\rm cm}^{-1}$.     

The organization of this paper is as follows. First, in Section \ref{FRIEDMANN EQUATIONS FOR SCALAR FIELD}, we present the Friedmann equations for a scalar field $\phi$. In Section \ref{matchingst} the internal closed FLRW metric has been matched with an external generalized Vaidya metric. The standard matching conditions are stated in this section.
Section \ref{BEHAVIOUR OF OSCILLATING SCALAR FIELD AS PRESSURE LESS FLUID} introduces the averaging procedure in the dark matter sector. We incorporate dark energy into our system in Section \ref{COLLAPSE DYNAMICS IN PRESENCE OF DARK ENERGY}. We used quintessence and phantom-like scalar fields to represent the dark energy. In Section \ref{Interaction with Dark energy} we present the main work of our paper which is related to the interaction of scalar field dark energy with dark matter in this collapsing system. The results of the paper are presented in Section \ref{Density Contrast}. The next section concludes the paper with a discussion.
\section{The dark matter scalar field and its mass}
\label{FRIEDMANN EQUATIONS FOR SCALAR FIELD}

Before we concentrate on the dark matter sector we give a glimpse of the model with which we will work. In this paper the total Lagrangian of the system is given by:
\begin{eqnarray}
\mathcal{S}=\int d^4x~\left(\mathcal{L}_{\text{grav}}+\mathcal{L}_\phi+\mathcal{L}_{\psi}+\mathcal{L}_{\text{int}}\right)\,\, ,
\end{eqnarray}
where the gravitational sector is given by the standard Einstein–Hilbert Lagrangian:
\begin{eqnarray}
\mathcal{L}_{\text{grav}}=\frac{\sqrt[•]{-g}R}{2}\,\, ,
\end{eqnarray} 
where $g$ is the determinant of the metric tensor $g_{\mu\nu}$ and $R$ is the Ricci scalar. The nature of the metric will be discussed in the next section. Here $\mathcal{L}_{\phi}$ is the Lagrangian for the $\phi$ field, which acts as the dark matter component of the universe. The other Lagrangian, $\mathcal{L}_{\psi}$,  is the Lagrangian of the quintessence or phantom-like scalar field, $\psi$, which is the dark energy component. One can also see that our model accommodates an explicit interaction term between the two kinds of fields whose Lagrangian is given by $\mathcal{L}_{\text{int}}$. This interaction is a non-minimal interaction in the dark sector. In the introduction we have cited some relevant works where we have non-minimal interaction in the dark sector. In most of those references, the dark energy sector was modeled by a quintessence or phantom-like scalar field, whereas the dark matter sector is modeled by a relativistic fluid with zero pressure. In this work, we assume both the sectors to be modeled by scalar fields which have a simple coupling as $q\phi^2 \psi^2$ where $q$ is the coupling constant. In \cite{Pettorino:2008ez}, the authors use a different kind of coupling at the level of fields representing the dark matter and dark energy sector.   We will introduce the dark energy sector Lagrangian and the interaction Lagrangian in later sections, in this section we primarily discuss the dark matter sector and the issue of the mass of the dark matter field. It turns out that the scalar field representing the dark matter candidate in our model is very light and can be termed as an axion-like particle.

In our work, we want to represent dark matter with a scalar field. We will start by considering a scalar field $\phi$ with Lagrangian density,
\begin{eqnarray}\label{Lagrangianeqpsi}
\mathcal{L}_{\phi}=-\sqrt[]{-g}~\left[\frac{1}{2}\partial_{\mu}\phi \partial^{\mu}\phi+V(\phi)\right]\ ,
\end{eqnarray}
where $V(\phi)$ is the potential of the scalar field. For the symmetry of the spherical collapse model, the scalar field will be a function of only time, $\phi=\phi(t)$. The scalar field is present both inside and outside of the spherical overdense region undergoing a gravitational collapse.  The spacetime of the spherical overdense region is modeled by a FLRW spacetime with spatial curvature constant $k$ and the scale-factor for the spacetime is assumed to be $a(t)$. The Hubble parameter for such a spacetime is $H\equiv \dot{a}(t)/a(t)$ where the dot specifies a derivative with respect to the coordinate time $t$.

The energy-momentum tensor for any scalar field or the interaction term with the Lagrangian $\mathcal{L}_X$  can be written as:
\begin{eqnarray}\label{EnergymomentumtensorFormula}
T^{(X)}_{\mu\nu}=\frac{-2}{\sqrt[]{-g}}\frac{\delta \mathcal{L}_{X}}{\delta g^{\mu\nu}}\,,
\end{eqnarray}
where $X$ can be either $\phi$, $\psi$ or ``int''.
Switching off the interaction term (for the time being), and varying the action with respect to $\phi$, yields the equation of motion for $\phi$:
\begin{eqnarray}\label{Kleinforpsi}
\ddot{\phi}+3H\dot{\phi}+ V'(\phi)=0,
\end{eqnarray}
where the prime represents a derivative with respect to the field. We will write the full equation of motion for $\phi$, including the non-minimal interaction term, later. 

In the entire work, we have written the equations for the expanding background spacetime and the collapsing spacetime separately as these dynamics differ. The exact equations for the background universe will have the scale-factor as $\bar{a}(t)$ and for the expanding universe, the spatial curvature constant, $k=0$. In contrast, we assume that we will have $k=1$ inside the overdense region. Here it must be pointed out that the scalar field both inside the collapsing patch and the expanding background are represented by the same field $\phi$ although they are two different functions of time. This is done to keep the notation straightforward because if we specify the field in the background by $\bar{\phi}$ one may think that it is a different scalar field outside. The spacetimes differ between the two regions of our concern and consequently, the same scalar field may behave differently in the two regions. We have only used the bars to specify the spacetime variables and the fluid variables (as energy density and pressure) in the background expanding universe.

As discussed in the introduction, we will model the dark matter sector by an oscillatory scalar field. This oscillatory scalar field will have a small mass parameter, $m$. The scalar field, representing the dark matter,  has a potential as $V(\phi)=\frac{1}{2}m^{2}\phi^{2}$. If we use this potential in Eq. (\ref{Kleinforpsi}), then the scalar field \(\phi\) will show oscillatory behavior when $m >>H$ \cite{Turner:1983he, Ratra:1990me}. This kind of scalar field can model a pressure-less fluid in an expanding universe. Of course, in our case, the non-minimal interaction term, in the dark sector, will back react on the oscillating dark matter sector. As long as this back reaction is not strong enough to unsettle the oscillations in the dark matter sector, we will be able to work with a coherently oscillating $\phi$ field. 

As briefly pointed out before, \( m \) does not carry the dimension of mass of the scalar field if we are not using natural units. In SI or CGS units, $m$,  has the unit of $L^{-1}$ where $L$ designates length. The parameter $m$ is expressed in units of energy in the natural unit system, where \( c = \hbar = 1 \). In geometrized units, $m$ is expressed in \( L^{-1} \). Similarly, \( H \) has the unit of \( T^{-1} \) in SI or CGS units, which can also be expressed in units of energy in natural units, and is expressed as \( L^{-1} \) in geometrized units. Here, we have taken \( m = 200\times 10^{-23} \, \text{cm}^{-1} \) and at time \( t = 0 \), using other parameter values, we get \( H = 2 \times 10^{-25}\, \text{cm}^{-1} \). We want to compare our values of \( m \) and \( H \) in units of energy. To convert these values to energy units, we use \( 1 \, \text{GeV} = 0.197\times 10^{-13} \, \text{cm}^{-1} \), resulting in \( m = 3.94 \times 10^{-26} \, \text{eV} \). For \(H\) at first we will change it into \( T^{-1} \) by multiplying \(c\) and using \( 1 \, \text{GeV} = 6.5826\times 10^{-25} \, \text{s}^{-1} \) we get, \( H = 3.9 \times 10^{-30} \, \text{eV} \). This gives a ratio of \( m/H \) around \( 10^4 \). Thus we ensure that $m \gg H$ in our case for the relevant time period of collapse.

Before we proceed further we introduce the external spacetime, the matching conditions, and the dynamics of the expanding background FLRW spacetime. In our case, the collapsing spacetime is matched with a generalized Vaidya spacetime. This matching occurs at a length scale smaller than the cosmological scale. In the cosmological scale dark energy dominates over dark matter producing an accelerated expansion of the universe. In this paper, we examine how the dark energy candidate influences the gravitational collapse of a spherical patch of dark matter at an intermediate cosmological scale. While dark matter dominates over dark energy at this scale, the effect of dark energy on the structure formation of dark matter cannot be neglected.

\section{Matching of spacetimes}
\label{matchingst}

In this paper, we want to model a spherically symmetric collapsing system. Here we have considered homogeneous and isotropic cosmology, and work with FLRW spacetime:
\begin{eqnarray} \label{FLRW}
ds^{2}=-dt^{2}+\frac{a^{2}(t)}{1-kr^{2}}dr^{2}+r^{2}a^{2}(t)(d\theta^{2}+\sin^{2}\theta d\psi^{2})\,\,,\nonumber\\ 
\end{eqnarray}
to model both the background universe and the over-dense region. Here $a(t)$ is the scale-factor of the overdense region and constant $k$ represents the curvature of the spatial section of the FLRW spacetimes, it can have values $0,~\pm1$. If $k=0$ then the spatial part will be flat whereas the negative and positive $k$ imply an open and closed spatial section. In our work, we want to generalize the top hat collapse using quintessence and phantom-like scalar field as dark energy and oscillating scalar field as dark matter. If we start with a spatially flat metric, the over-dense region will never reach a turn-around radius. Therefore we have considered  $k=1$, closed FLRW metric to model the over-dense region and the background will be modeled by spatially flat FLRW metric:
\begin{eqnarray} \label{FLRWBACKGROUND}
ds^{2}=-dt^{2}+\bar{a}^{2}(t)dr^{2}+r^{2}\bar{a}^{2}(t)(d\theta^{2}+\sin^{2}\theta d\psi^{2})\,\,,\nonumber\\ 
\end{eqnarray}
where $\bar{a}(t)$ is the scale-factor for background universe.

The dynamic evolution of the system will be determined by the Einstein equations,
which for the metric given in Eq.(\ref{FLRW}), can be written as
\begin{align}
    \rho &= \frac{F'}{r^{2}a^{3}}, & P &= \frac{-\dot{F}}{r^{3}a^{2}\dot{a}}\,,
\end{align}
where $F(t,r)$ is the Misner-Sharp mass term which is interpreted as the energy density inside a spherical shell of collapsing matter at a given comoving time. Here it will be:
\begin{eqnarray}\label{Fform}
    F(r)=(\dot{a}^{2}+k)ar^{3}\,.
\end{eqnarray}
The prime specifies a derivative with respect to $r$ and the dot specifies a time derivative. Here $r$ is a co-moving coordinate and does not change with time.

In a gravitational collapse involving dark matter and dark energy constituents, the dark energy constituent may not exactly follow the collapsing dark matter cloud. Consequently, there may be finite pressure on the surface of the spherical patch due to the dark energy constituent. As a result of this as the system starts to collapse there will be a matter flux from the boundary of the over-dense region. Here our system is comprised of an interior spacetime with a closed FLRW metric and a background spacetime with a flat FLRW metric. However, to represent the matter flux through the boundary of the overdense region we have considered the immediate neighborhood of the overdense region as a generalized Vaidya spacetime. 

As we mentioned above, we consider a spacetime configuration that is internally ($\mathcal{V}^{-}$) a closed FLRW spacetime and externally ($\mathcal{V}^{+}$) an exploding generalized Vaidya spacetime :
\begin{eqnarray}
dS^2_{-}&=&-dt^2+a^2(t)\left(\frac{dr^2}{1-r^2}+r^2d\Omega^2\right)\,\,\nonumber\\
&=& -dt^2 + a^2(t) d\Phi^2 +a^2(t)\sin^2\Phi d\Omega^2\,\, ,\\
\nonumber\\
dS^2_{+}&=& -\left(1-\frac{2M(r_v , v)}{r_v}\right)dv^2 - 2dv dr_v + r_v^2 d\Omega^2\,\, , \nonumber\\
\end{eqnarray}
where we consider co-moving radius $r=\sin\Phi$.  Here $r_v$ and $v$ are the coordinates and $M(r_v , v)$ is the mass function, all of which are used to specify the generalized Vaidya spacetime.
At the boundary of the timelike hypersurface ($\Sigma$) $r=r_{b}$, the internal closed FLRW spacetime smoothly matches the external generalized Vaidya spacetime. 

Since the flux of energy emanating from the surface of the spherical patch is generally not expected to behave like null dust, it can be modeled as a combination of null dust and matter. Accordingly, we consider the generalized Vaidya spacetime as the external spacetime surrounding the internal closed FLRW spacetime. The stress-energy tensor ($T_{\mu\nu}$) of the generalized Vaidya spacetime can thus be expressed as the sum of the stress-energy tensor of null dust ($T^{(n)}_{\mu\nu}$) and the stress-energy tensor of matter ($T^{(m)}_{\mu\nu}$) \cite{Mken}:
\begin{eqnarray}
T_{\mu\nu} = T^{(n)}_{\mu\nu} + T^{(m)}_{\mu\nu}, 
\end{eqnarray}
where $T^{(n)}_{\mu\nu} = \rho_n l_{\mu}l_{\nu}$ and $$T^{(m)}_{\mu\nu} = (\mu + \varrho)(l_{\mu}n_{\nu} + l_{\nu}n_{\mu}) + \varrho g_{\mu\nu}\,,$$ with $l_\mu$ and $n_\mu$ being the outgoing and ingoing null vectors, respectively, satisfying the conditions: $l^\mu l_\mu = n^\mu n_\mu = 0$ and $l^\mu n_\mu = -1$. Therefore, the four-dimensional spacetime manifold with $SO(3)$ symmetry can be written as $\mathcal{M} = l \times n \times S^2$.
One can show for the above metric of generalized Vaidya spacetime:
\begin{eqnarray}
\rho_n = \frac{2\dot{M}(r_v, v)}{r^2}, \mu=\frac{2M'(r_v, v)}{r^2}, \varrho = -\frac{M''(r_v, v)}{r},
\end{eqnarray}
where
\begin{equation}
\dot{M}(r_v,v)\equiv \frac{\partial M(r_v,v)}{\partial v}, \, M'(r_v,v)\equiv  \frac{\partial M(r_v,v)}{\partial r_v}\,.
\end{equation}
At the timelike hyper-surface ($\Sigma$) the internal and external spacetimes match with each other, $\Phi=\Phi_b$ and the $v$ and $r_v$ become the function of co-moving time $t$. Therefore, at $\Sigma$, we can write down the induced metric from both sides as,
\begin{eqnarray}
dS^2_{-}|_{\Sigma}&=& -dt^2 + a^2(t)\sin^2\Phi_b d\Omega^2\,\, ,\\
\nonumber\\
dS^2_{+}|_{\Sigma}&=& -\left(\dot{v}^2-\frac{2M(r_v , v)}{r_v}\dot{v}^2+2\dot{v}\dot{r}_v\right)dt^2+ r_v^2 d\Omega^2\,\, , \nonumber\\
\end{eqnarray}
where $\dot{v}$ and $\dot{r}_v$ are the partial derivatives of $v$ and $r_v$ with respect  to coordinate time $t$. 
For the smooth matching of two spacetimes at a hypersurface, the necessary and sufficient condition is that the induced metric ($h_{ab}$) and the extrinsic curvature ($K_{ab}$) from both sides should match at the junction. Matching the induced metric we get:
\begin{eqnarray}\label{inducedmatchingcondition}
\left(\dot{v}^2-\frac{2M(r_v , v)}{r_v}\dot{v}^2+2\dot{v}\dot{r}_v\right) &=& 1\,\, ,\\
r_v &=& a(t)\sin\Phi_b\,\, .
\end{eqnarray}
The first fundamental forms match on the timelike hypersurface $\Sigma$ when the above condition is met. 

Next, to match the second fundamental form (extrinsic
curvatures) for the interior and exterior metrics, we note
that the normal to the hypersurface $\Sigma$, as calculated from
the interior FLRW metric is given as,
$$n^{\alpha}_{-}\equiv \lbrace 0,a(t)^{-1},0,0\rbrace \,,$$
and the normal as derived
from the generalized Vaidya space-time is
\begin{eqnarray}
    n^{\alpha}_{+}&\equiv& \lbrace-\frac{1}{\sqrt{1-\frac{2M}{r_v}+2\frac{dr_v}{dv}}}, \frac{1-\frac{2M}{r_v}+\frac{dr_v}{dv}}{\sqrt{1-\frac{2M}{r_v}+2\frac{dr_v}{dv}}},0,0\rbrace\,\, .\nonumber\\
\end{eqnarray}
The extrinsic curvature of the hypersurface is defined as,
\begin{eqnarray}
    K_{ab}=\frac{1}{2}[g_{ab,c}n^{c}+g_{cb}n^{c}_{,a}+g_{ac}n^{c}_{,b}]\,,
\end{eqnarray}
where commas specify partial derivatives with respect to coordinate variables.
Setting $[K_{\theta\theta}^{-}-K_{\theta\theta}^{+}]_{\Sigma}=0$ on the hypersurface $\Sigma$, we get:
\begin{equation}\label{Extrinsiccurvaturetheta}
\cos\Phi_b = \frac{1-\frac{2M}{r_v}+\frac{dr_v}{dv}}{\sqrt{1-\frac{2M}{r_v}+2\frac{dr_v}{dv}}}\,.
\end{equation}
Using above equations and Eq. (\ref{Fform}) we obtain:
\begin{equation}
F(t,\sin\Phi_b)=2M(r_v,v)\,,
\end{equation}
where $F$ is the Misner-Sharp mass of the internal collapsing spacetime.
Equating the temporal components ($K_{tt}$ from both sides) on the hypersurface we get:
\begin{equation}
M(r_v,v)_{,r_v}=\frac{F}{2\sin\Phi_b a(t)}+\sin^2\Phi_b a\ddot{a}\,\ .
\end{equation}
These are all the relevant conditions for matching the internal collapsing FLRW spacetime and the generalized Vaidya spacetime surrounding the collapsing patch.
\section{Oscillating scalar field mimicking a pressure less fluid}
\label{BEHAVIOUR OF OSCILLATING SCALAR FIELD AS PRESSURE LESS FLUID}

To proceed from this point we will specifically write the oscillating field $\phi$ in terms of functions of $t$ as
\begin{equation}
\phi(t)=\phi_{+}(t) \sin \alpha(t)+\phi_{-}(t) \cos \alpha(t),
\end{equation}
where $\alpha(t)$ and $\phi_{ \pm}(t)$ are two time-varying functions where it is assumed that the rate of change of $\alpha(t)$ is much greater than the rate of change of $\phi_{\pm}(t)$. In this section we follow the conventions used in Ref.~\cite{Ratra:1990me}.  We will soon find out the form of these functions.
Fixing  the form of the potential as $V(\phi)=\frac{1}{2}m^2\phi^{2}$ in Eq.(\ref{Kleinforpsi})  we get 
multiple conditions which help us to find out the unknown functions $\alpha(t), \phi_{\pm}(t)$.
The differential equation for $\alpha(t)$ is:
\begin{equation}
\dot{\alpha}^{2}-m^{2}=0.
\label{m2term}
\end{equation}
This equation can be solved to find out $\alpha$ as we have $\dot{\alpha}=m$ (taking the positive square root) whose solution is $\alpha(t)=m\left(t-t_{0}\right)$ where we can set $t_{0}=0$. Using this solution we get the following differential equation of $\phi_{\pm}(t)$:
\begin{equation}\label{psidotequ}
\dot{\phi}_{ \pm}+\frac{3}{2} H \phi_{ \pm}=0\,.
\end{equation}
The solution of this differential equation is of the form:
\begin{equation}\label{phipm}
\phi_{ \pm}(t)=\frac{C_{ \pm}}{a(t)^{3 / 2}}\,,
\end{equation}
where $C_{ \pm}$ are dimensionless constants. It is seen that $\phi_{\pm}(t)$ changes in the cosmological time scale, whereas $\alpha(t)$ changes at a time rate of $m$. As $m\gg H$,  $\alpha(t)$ changes much faster than $\phi_{\pm}(t)$. This fact gives us the freedom to average out the fast modes.
The oscillating scalar field will act as the dark matter candidate in the expanding background spacetime as well as in the collapsing patch. In the background expansion case, the scale-factor will be represented by $\bar{a}$, and similarly, the Hubble parameter will be $\bar{H}$. In this section we specify the results in the overdene spherical patch. Because the theory, as presented in this section, is essentially the same (for the dynamics of the background and the overdense patch) we do not again present the results in terms of the barred variables for the case involving background expansion, the reader will easily be able to understand from the context when the barred variables will be used and where unbarred variables become useful.

To obtain the time-averaged form of the Friedmann equations, it can be written as follows:
\begin{eqnarray}\label{Friedmanneqforpsiave}
\frac{3\dot{a}^2}{a^2}+\frac{3k}{a^2}=\left\langle\frac{1}{2}\dot\phi^2+V(\phi)\right\rangle\, ,
\end{eqnarray}
\begin{eqnarray}\label{Friedmanneqforpsiave}
-\frac{2\ddot{a}}{a}-\frac{\dot{a}^2}{a^2}-\frac{k}{a^2}=\left\langle\frac{1}{2}\dot\phi^2-V(\phi)\right\rangle \, ,
\end{eqnarray}
here the angular brackets specify the time average. The average of the product of two functions of time, as $\langle s(t) f(t)\rangle$, where $s(t)$ varies much slower in time, compared to one period of oscillation for the fast oscillating function $f(t)$, can be written as $\langle s(t) f(t)\rangle=s(t)\langle f(t)\rangle$. If $f(t)$ has a time period $T=2 \pi / m$ then,
\begin{equation}
\langle f(t)\rangle=\frac{m}{2 \pi} \int_{0}^{\frac{2 \pi}{m}} f\left(t^{\prime}\right) d t^{\prime} .
\end{equation}
In our case $\sin \alpha, \cos \alpha, \sin ^{2} \alpha, \cos ^{2} \alpha$ are all oscillating functions with period $2 \pi / m$. Using this we will get,
\begin{equation}
\left\langle\dot{\phi}^{2}\right\rangle=\frac{1}{2}\left[\dot{\phi}_{+}^{2}+\dot{\phi}_{-}^{2}+m^{2}\left(\phi_{+}^{2}+\phi_{-}^{2}\right)\right]+m\left(\dot{\phi}_{-} \phi_{+}-\dot{\phi}_{+} \phi_{-}\right)\,,
\end{equation}
and
\begin{equation}\label{phisqeq}
\left\langle\phi^{2}\right\rangle=\frac{1}{2}\left(\phi_{+}^{2}+\phi_{-}^{2}\right).
\end{equation}
We can now evaluate the system's time-averaged energy density and pressure using these values.

The energy density of the system is:
\begin{equation}\label{rhoforpsi}
\langle\rho_{\phi}\rangle = \frac{1}{2}\left\langle\dot{\phi}^{2}+m^{2} \phi^{2}\right\rangle=\frac{1}{2} m^{2}\left(C_{+}^{2}+C_{-}^{2}\right) \frac{1}{a^{3}}\left(1+\frac{9}{8} \frac{H^{2}}{m^{2}}\right).
\end{equation}
The pressure of the system is given by,
\begin{equation}
\langle P_{\phi}\rangle = \frac{1}{2}\left\langle\dot{\phi}^{2}-m^{2} \phi^{2}\right\rangle=\frac{9}{16} m^{2}\left(C_{+}^{2}+C_{-}^{2}\right) \frac{1}{a^{3}}\left(\frac{H^{2}}{m^{2}}\right).
\end{equation}
The ratio of the pressure and the energy density of the system yields:
\begin{equation*}
\frac{\langle P_{\phi}\rangle}{\langle\rho_{\phi}\rangle} =\frac{\frac{9}{8} \frac{H^{2}}{m^{2}}}{\left(1+\frac{9}{8} \frac{H^{2}}{m^{2}}\right)}. \tag{23}
\end{equation*}
In the limit $m \gg H$ we see that the above ratio tends to zero, specifying an effective equation of state (EoS) of dust. 

The results of this section exactly hold when we have only the $\phi$ field in the universe and $k=0$ and approximately hold when $k\ne 0$ and we have multicomponent matter sources. We have checked that the above results in general hold excellently if initially, the dust energy density exceeds all other forms of energy density. In a dust dominated universe, where other sources of energy density are present, the other matter or curvature energy density components do have a negligible effect on the initial conditions in the $\phi$ sector (dust sector) and hence all the above relations work approximately well. If the dark sector non-minimal coupling term remains small the above approximation scheme even works at later stages. In this paper, we have used these facts to find the dynamics of gravitational collapse, in presence of non-minimal coupling in the dark sector, where the collapse happens in a dust dominated universe.

The above analysis shows that the scalar field behaves like dust when $m \gg H$. The equivalence of the two systems is established when we integrate the fast-oscillating degree of freedom in the scalar field solution. In cosmology, we generally do not care about the fast changes, compared to the time scale $1 / H$, therefore in this limit, we can assume that the scalar field system, in the presence of the harmonic potential, behaves like dust.

\begin{figure*}
\centering
\begin{minipage}[t]{.4\textwidth}
\centering
\includegraphics[width=70mm,height=52mm]{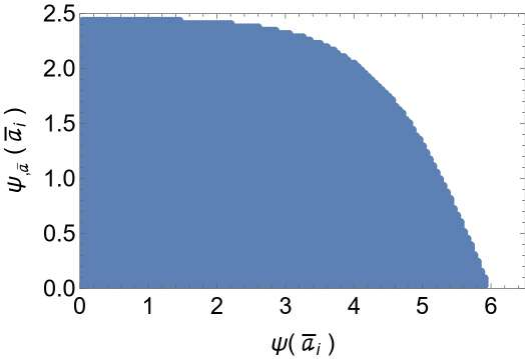}
\end{minipage}\qquad
\begin{minipage}[t]{.4\textwidth}
\centering
\includegraphics[width=70mm,height=52mm]{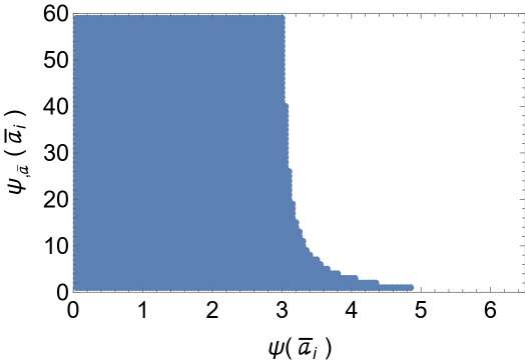}
\end{minipage}
\caption{ The region plots show the relationship between the initial values of the background dark energy scalar field. $\psi(\bar{a})$, and its derivative with respect to the background scale-factor, $\psi_{,\bar{a}}(\bar{a})$, are both evaluated at $\bar{a_{i}}(t) = 1.0003/u$. The dark energy scalar field, $\psi(\bar{a})$, is dimensionless, while its derivative is expressed in units of $u$, where $\bar{a}$ represents the background scale-factor. The left panel illustrates the quintessence dark energy model and the right panel corresponds to the phantom dark energy model. The parameter values are set as follows: $q = 0.01 u^{2}$, $V_{0} = 0.1 u^{2}$, $\lambda = 1$, $m = 2 \times 10^{4} u$, and $C_{\pm} = 2 \times 10^{-4}/u^{3/2}$, with $u = 10^{-25} \text{cm}^{-1}$. The shaded regions in both plots represent the range of initial values for which the scalar fields drive sufficient acceleration in the background universe for the quintessence (left) and phantom (right) models.}
\label{phiprimevsphiplotbackground}
\end{figure*}

\begin{figure*}
\centering
\begin{minipage}[t]{.4\textwidth}
\centering
\includegraphics[width=70mm,height=52mm]{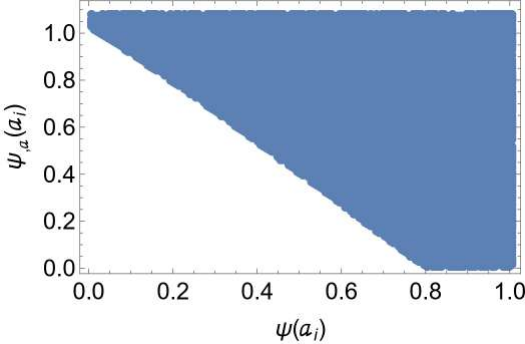}
\end{minipage}\qquad
\begin{minipage}[t]{.4\textwidth}
\centering
\includegraphics[width=70mm,height=52mm]{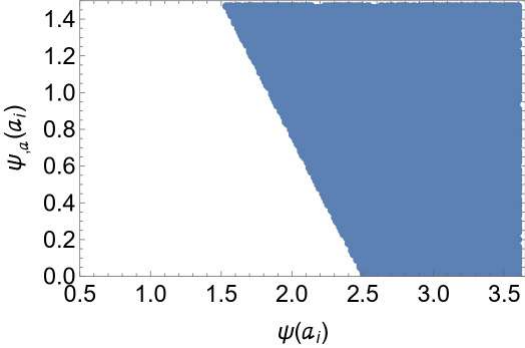}
\end{minipage}
\caption{ The region plots show the relationship between the initial values of the dark energy scalar field present inside the overdense spherical patch. $\psi(a)$, and its derivative with respect to the scale-factor inside the patch, $\psi_{, a}(a)$, both are evaluated at $a_{i}(t) = 1/u$. The dark energy scalar field, $\psi(a)$, is dimensionless, while its derivative is expressed in units of $u$, where $a$ represents the scale-factor inside the patch. The left panel illustrates the quintessence dark energy model and the right panel corresponds to the phantom dark energy model. The parameter values are set as follows: $q = 0.01 u^{2}$, $V_{0} = 0.1 u^{2}$, $\lambda = 1$, $m = 2 \times 10^{4} u$, and $C_{\pm} = 2 \times 10^{-4}/u^{3/2}$, with $u = 10^{-25} , \text{cm}^{-1}$. The shaded regions in the plots represent the range of initial values within the inner over-dense region where the system, which includes either the quintessence (left) or phantom (right) fields, will undergo collapse.}
\label{phiprimevsphiplotinside}
\end{figure*}
\section{Collapse dynamics in the presence of dark energy}
\label{COLLAPSE DYNAMICS IN PRESENCE OF DARK ENERGY}

In the previous section it was established that the scalar field $\phi(t)$ can behave like dust.  The other scalar field $\psi=\psi(t)$ will be considered as the dark energy constituent. The Lagrangian density of this scalar field is given by,
\begin{eqnarray}\label{Lagrangianeqphi}
\mathcal{L}_{\psi}=-\sqrt[]{-g}~\left[\frac{1}{2}\epsilon\partial_{\mu}\psi \partial^{\mu}\psi+V(\psi)\right]\, ,
\end{eqnarray}
where $V(\psi)$ is the potential of the scalar field and $\epsilon=1 (-1)$ denotes quintessence (phantom) field respectively. 
Using Eq.~(\ref{EnergymomentumtensorFormula}) the energy-momentum tensor for this scalar field will be,
\begin{eqnarray}\label{Energymomentumtensorforscalarfieldphi}
T^{(\psi)}_{\mu\nu}=\epsilon\partial_{\mu}\psi \partial_{\nu}\psi-g_{\mu\nu}\left[\frac{1}{2}\epsilon\partial_{\mu}\psi \partial^{\mu}\psi+V(\psi)\right]\, .
\end{eqnarray}
For a system containing both the scalar fields $\psi(t)$ and $\phi(t)$ the Friedmann equations become:
\begin{eqnarray}\label{Friedmanneqt1}
\frac{3\dot{a}^2}{a^2}+\frac{3k}{a^2}=\rho_{\phi}+\rho_{\psi}\,\, ,
\end{eqnarray}
\begin{eqnarray}\label{Friedmanneqt2}
-\frac{2\ddot{a}}{a}-\frac{\dot{a}^2}{a^2}-\frac{k}{a^2}= P_{\phi} +P_{\psi}\,\, ,
\end{eqnarray}
here,  $\rho_{\psi} = \frac{\epsilon\dot{\psi}^2}{2} + V_0e^{-\lambda\psi}$ and $P_{\psi} = \frac{\epsilon\dot{\psi}^2}{2} - V_0e^{-\lambda\psi}$, are the relevant energy density and pressure terms for scalar field $\psi(t)$ calculated using Eq. (\ref{Energymomentumtensorforscalarfieldphi}), and 
\begin{eqnarray}\label{potentialpsi}
V(\psi)=V_{0}e^{-\lambda \psi}\,,
\label{potf}
\end{eqnarray}
where $\lambda$ is a dimensionless constant. The above equations do not contain the time-averaged form. Later when we include time averaging over the fast degrees of freedom in the $\phi$ sector we will have the time average pressure $\left\langle P_{\phi}\right\rangle$ for the oscillating scalar to be approximately zero and time average energy density to be approximately $\left\langle\rho_{\phi}\right\rangle \propto a^{-3}$. The approximations are there because ideally the above mentioned values will differ from their idealized values (which happens when there is only a single field $\phi$ oscillating coherently) when dark energy and its interactions are taken into account. In this paper we have shown that the coherent oscillation model for the dark matter sector remains workable when there is an interaction term. The point will be elaborately discussed in the next subsection.  In absence of any non-minimal coupling, the field equations for these two scalar fields will be,
\begin{eqnarray}\label{Klein4}
   \ddot{\phi} + 3H\dot{\phi} + m^2\phi = 0 ,
\end{eqnarray}
\begin{eqnarray}\label{Klein5}
   \epsilon \ddot{\psi}+3H\epsilon\dot{\psi}-\lambda V_{0}e^{-\lambda \psi}=0 .
\end{eqnarray}
We do not assume any fast degree of freedom inside the expression of $\psi(t)$ and consequently its value changes in the cosmological time scale. Next, we introduce a possible (non-minimal) interaction between the dark matter field and the dark energy field.

The results in this section focused on the dynamics of the collapsing patch, for the background expansion similar equations hold, the scale factor, energy density pressure, and the Hubble parameter have to be replaced by the barred analogs.
\begin{figure*}
\subfigure[Evolution of Hubble parameter H over time within the overdense region.]
{\includegraphics[width=81mm,height=55mm]{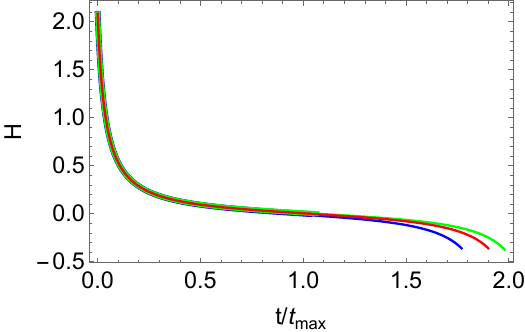}\label{1a}}
\hspace{0.2cm}
\subfigure[Evolution of relative Scale-factor over time within the overdense region.]
{\includegraphics[width=81mm,height=55mm]{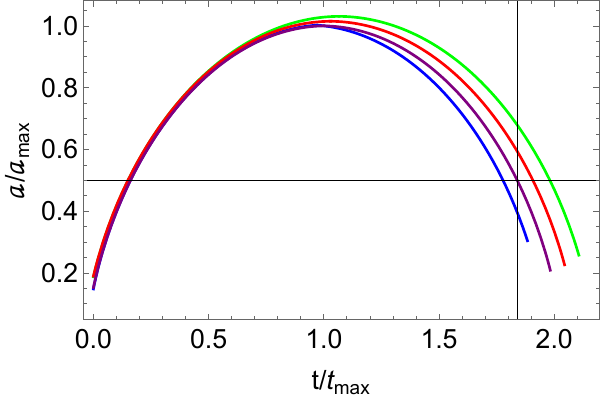}\label{1b}}
\hspace{0.2cm}
\subfigure[Variation of the ratio of dark energy densities over time between the overdense region $\rho_{\psi}$ and the background universe $\bar{\rho}_{\psi}$.]
{\includegraphics[width=81mm,height=55mm]{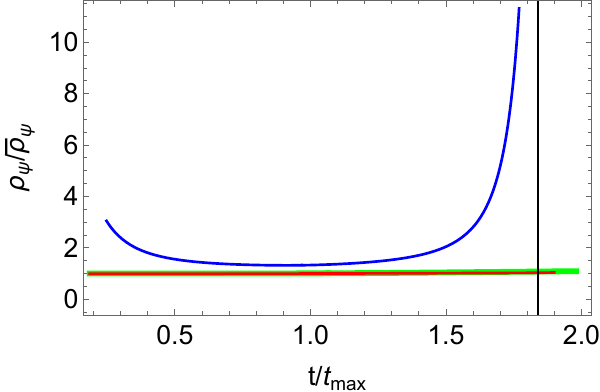}\label{1c}}
\hspace{0.2cm}
\subfigure[Variation of the dark energy equation of state $\omega_{\psi}$ in the overdense region over time.]
{\includegraphics[width=81mm,height=55mm]{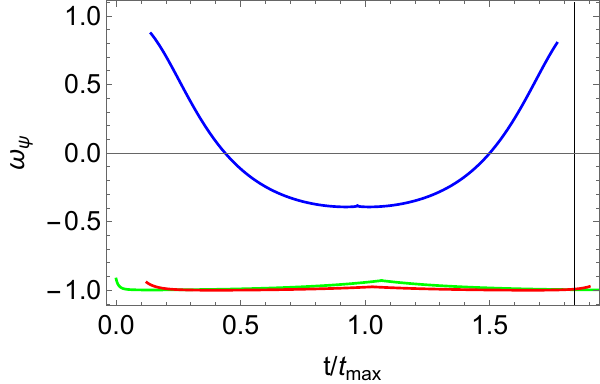}\label{1d}}
\hspace{0.2cm}
\subfigure[Variation of the total equation of state $\omega_{t}$ in the overdense region over time.]
{\includegraphics[width=81mm,height=55mm]{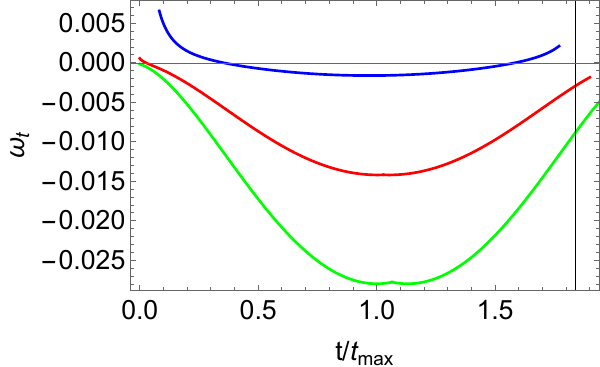}\label{1e}}
\hspace{0.2cm}
\subfigure[Variation of the total equation of state $\bar{\omega}_{t}$ over time for the background universe.]
{\includegraphics[width=81mm,height=55mm]{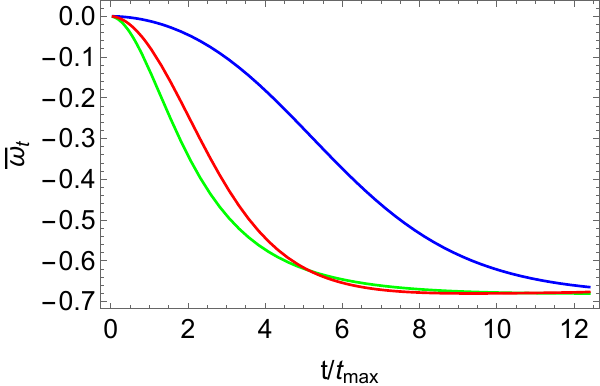}\label{1f}}
\hspace{0.2cm}
\caption{Time evolution of various variables for different scenarios with two coupling constants have been shown. The scalar field for the dark energy is modeled as a quintessence field with a potential $V=V_{0}e^{-\lambda \psi}$, where $\lambda=1$. The mass for the scalar field dark matter $m= 2\times 10^{4} u$ and the Hubble parameter at initial time $t_{i}$ is $H(t_{i})=2u$. The purple colored curve represents the standard top-hat collapse scenario (in the absence of any dark energy candidate $\psi$). For a fixed value of coupling constant $q=0.01u^{2}$, the red and green colored curves correspond to dark energy potentials with $V_{0}=0.05u^{2}$ and $V_{0}=0.1u^{2}$, respectively. The blue colored curve corresponds to a different coupling constant $q=5000u^{2}$ and $V_{0}=0.1u^{2}$. Here, $a_{\text{max}}$ denotes the value of the scale-factor at the turnaround point within the overdense region during the top-hat collapse, while $t_{\text{max}}$ represents the time at which the system reaches this turnaround point for top hat collapse and $u=10^{-25} \text{cm}^{-1}$.}
\label{Time evolution for quintessence field}
\end{figure*}
\subsection{Interaction with dark energy}
\label{Interaction with Dark energy}

Now we will incorporate an interaction between these two scalar fields with an interaction Lagrangian as,
\begin{eqnarray}
\mathcal{L}_{\text{int}} &=& -\sqrt[]{-g}[q\psi^{2}\phi^{2}]\,,
\end{eqnarray}
where $q$ is a coupling constant with an unit of $L^{-2}$ where $L$ specifies length. If we vary this Lagrangian density with respect to the metric we get
$$\delta \mathcal{L}_{\text{int}} = \frac{1}{2} \sqrt[]{-g} g_{\mu\nu}\delta g^{\mu\nu} q\psi^{2}\phi^{2}\,.$$
Using the above equation and Eq.~(\ref{EnergymomentumtensorFormula}) we get,
\begin{eqnarray}\label{EnergymomentumtensorforInteraction}
T^{(\text{int})}_{\mu\nu}=- q g_{\mu\nu}\psi^{2}\phi^{2}\,.
\end{eqnarray}
From Eq.~(\ref{EnergymomentumtensorforInteraction}) the energy-momentum tensor components will be,
\begin{eqnarray}\label{rhoint}
    -T^{0 (\text{int})}_{0}=q\psi^{2}\phi^{2}=\rho_{\text{int}}
\end{eqnarray}
\begin{eqnarray}\label{pint}
    T^{i (\text{int})}_{i}=-q\psi^{2}\phi^{2}=P_{\text{int}}.
\end{eqnarray}
Therefore with this new interaction term, the total energy-momentum tensor is
\begin{eqnarray}\label{Energymomentumtensortotal}
  T^{(\text{tot})}_{\mu\nu}=T^{(\phi)}_{\mu\nu}+T^{(\psi)}_{\mu\nu}+T^{(\text{int})}_{\mu\nu}\,.
\end{eqnarray}
In our formalism, the interaction between the two scalar fields introduces a new energy-momentum tensor.

Finally, the Friedmann equations for the total system can be written as
\begin{eqnarray}\label{Friedmanneqtplus1}
\frac{3\dot{a}^2}{a^2}+\frac{3k}{a^2}=\rho_{\phi} +\rho_{\psi}+\rho_{\text{int}}\,\, ,
\end{eqnarray}
\begin{eqnarray}\label{Friedmanneqtplus2}
-\frac{2\ddot{a}}{a}-\frac{\dot{a}^2}{a^2}-\frac{k}{a^2}= P_{\phi} + P_{\psi}+ P_{\text{int}}\,\, ,
\end{eqnarray}
where we have not executed a time averaging in the $\phi$ sector, these are the standard Friedmann equations.
The modified field equations for the two scalar fields will be,
\begin{eqnarray}\label{Kleinfortot1}
\ddot{\phi}+3H\dot{\phi}+(m^{2}+2q\psi^{2})\phi=0,
\label{sfein}
\end{eqnarray}
\begin{eqnarray}\label{Kleinfortot2}
\epsilon\ddot{\psi}+3H\epsilon\dot{\psi}-\lambda V_{0} e^{-\lambda \psi}+2q \phi^{2}\psi=0\,.
\end{eqnarray}
The above set of equations are a generalization of the top-hat collapse model. If we assume a closed FLRW universe where the relevant fluid variables are $\rho_m$ (energy density) and $p_m$ (pressure) then the fundamental equations in the top-hat collapse model are given by \cite{Gunn:1972sv}:
\begin{eqnarray}
\rho_{m}&=&\frac{3\dot{a}^{2}}{a^{2}}+\frac{3k}{a^{2}}\,\, \label{one},\\
p_{m}&=&-\frac{2\ddot{a}}{a}-\frac{\dot{a}^{2}}{a^{2}}-\frac{k}{a^{2}}\,, 
\label{two}
\end{eqnarray}
along with the continuity equation for matter:
\begin{eqnarray}
    \dot{\rho}_m + 3\frac{\dot{a}}{a}(\rho_m+p_{m}) = 0\,\, .
\label{three}
\end{eqnarray}
In the standard top-hat model the Friedmann equations can be solved parametrically when we assume that we have pressure-less dark matter like candidate in the universe. In the top-hat model $p_m=0$. If one wants to numerically solve the system then one may use  Eq.~(\ref{one}) in conjunction with Eq.~(\ref{three}) to solve the dynamics of the system. In the present case, we generalize the top-hat collapse to the case where the dark matter sector is accompanied by a non-minimally coupled dark energy sector in the universe. Both the dark matter and dark energy sectors are modeled by scalar fields and their non-minimal interaction is modeled by scalar field interaction. In such a case Eq.~(\ref{one}) and  Eq.~(\ref{two}) corresponds to Eq.~(\ref{Friedmanneqtplus1}) and Eq.~(\ref{Friedmanneqtplus2}) respectively. The energy-momentum conservation condition (or the continuity equation), given in Eq.~(\ref{three}), generalizes to the field equations for $\phi$ and $\psi$ in Eq.~(\ref{Kleinfortot1}) and Eq.~(\ref{Kleinfortot2}) respectively.  In this way the present paper attempts to generalize the top-hat collapse model so that the general model shows how gravitational collapse happens in an interacting dark sector.

Before calculating the dynamics of the system we need to keep in mind that the two Friedmann equations and the two scalar field equations are not independent of each other; only three of them are independent. In the next subsection, we will integrate out the fast degrees of freedom in the $\phi$ sector and work with the time-averaged forms of $\rho_\phi$ and $P_\phi$. One must note that $\phi(t)$ will behave like dust only when the mass term, $m$, is large compared to $H$. Henceforth we will try to maintain this condition and determine how the collapse dynamics proceeds.
\begin{figure*}
\subfigure[Evolution of Hubble parameter H over time within the over-dense region.]
{\includegraphics[width=81mm,height=55mm]{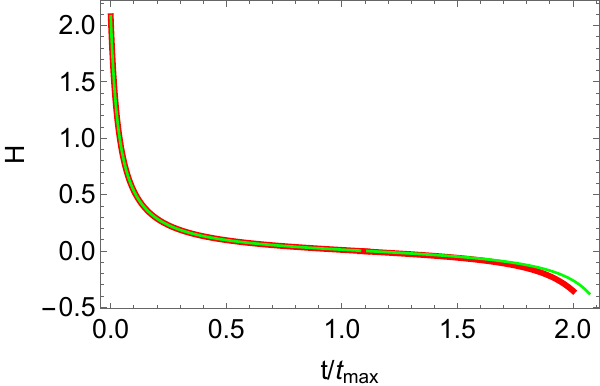}\label{2a}}
\hspace{0.2cm}
\subfigure[Variation of relative Scale-factor over time within the overdense region.]
{\includegraphics[width=81mm,height=55mm]{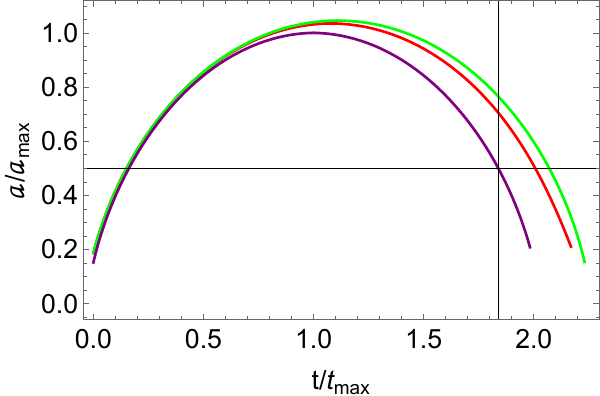}\label{2b}}
\hspace{0.2cm}
\subfigure[Variation of the ratio of dark energy densities over time between the overdense region $\rho_{\psi}$ and the background universe $\bar{\rho}_{\psi}$.]
{\includegraphics[width=81mm,height=55mm]{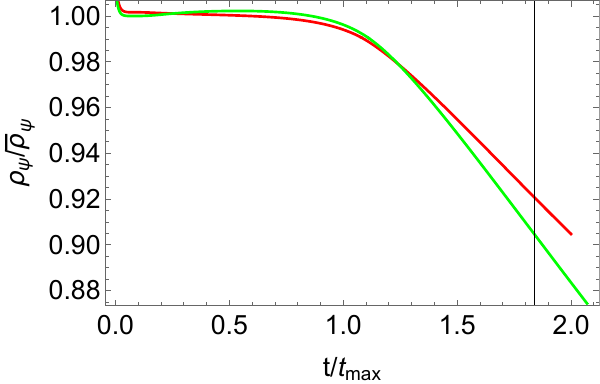}\label{2c}}
\hspace{0.2cm}
\subfigure[Variation of the dark energy equation of state $\omega_{\psi}$ in the overdense region over time.]
{\includegraphics[width=81mm,height=55mm]{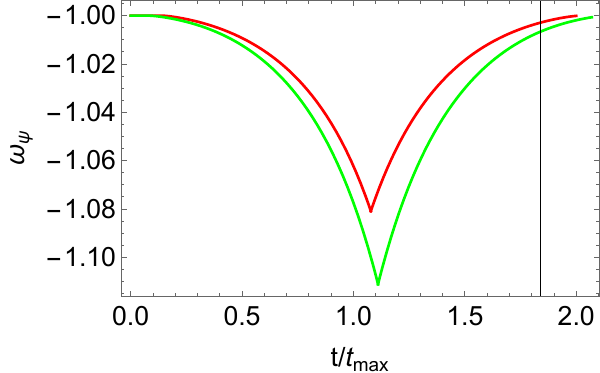}\label{2d}}
\hspace{0.2cm}
\subfigure[Variation of the total equation of state $\omega_{t}$ in the overdense region over time.]
{\includegraphics[width=81mm,height=55mm]{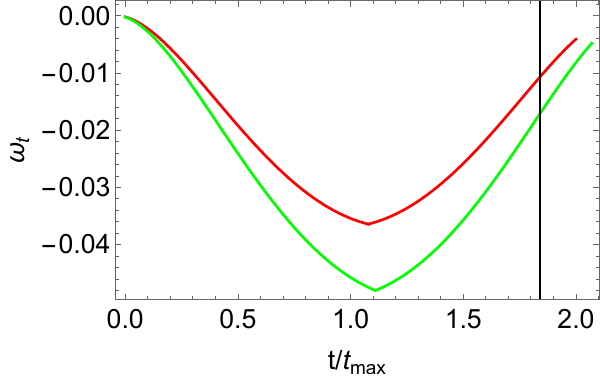}\label{2e}}
\hspace{0.2cm}
\subfigure[Variation of the total equation of state $\bar{\omega}_{t}$ over time for the background universe.]
{\includegraphics[width=81mm,height=55mm]{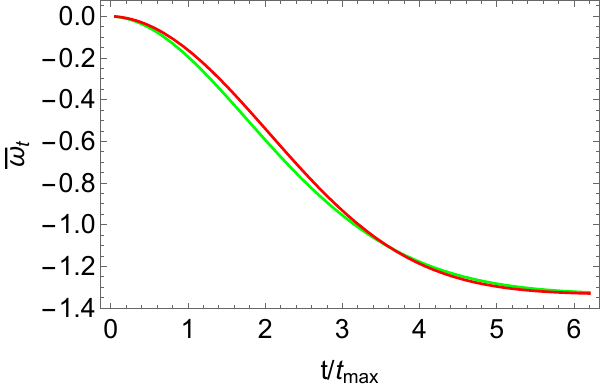}\label{2f}}
\hspace{0.2cm}
\caption{Time evolution of various variables for different scenarios with a coupling constant $q=0.01u^{2}$ have been shown. The scalar field for the dark energy is modeled as a Phantom field with a potential $V=V_{0}e^{-\lambda \psi}$, where $\lambda=1$. The mass for the scalar field dark matter $m= 2\times 10^{4} u$ and the Hubble parameter at initial time $t_{i}$ is $H(t_{i})=2u$. The purple colored curve represents the standard top-hat collapse scenario (in the absence of any dark energy candidate $\psi$). The red and green colored curves correspond to dark energy potentials with $V_{0}=0.08u^{2}$ and $V_{0}=0.1u^{2}$, respectively. Here, $a_{\text{max}}$ denotes the value of the scale-factor at the turnaround point within the over-dense region during the top-hat collapse, while $t_{\text{max}}$ represents the time when the system reaches this turnaround point for top hat collapse and $u=10^{-25} \text{cm}^{-1}$.}
\label{Time evolution for phantom field}
\end{figure*}
\begin{figure*}
\subfigure[]
{\includegraphics[width=70mm,height=52mm]{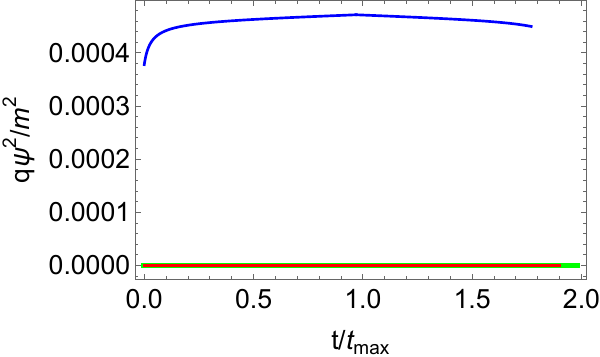}\label{qpsisqmsqinside}}
\subfigure[]
{\includegraphics[width=70mm,height=52mm]{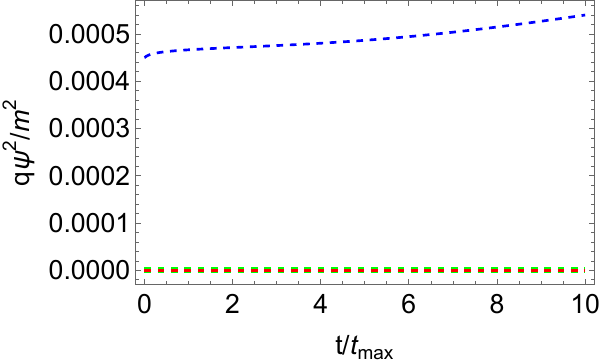}\label{qpsisqmsqbackground}}
\caption{Time evolution of the quantity $q\psi^{2}/m^{2}$ for the quintessence field with a potential $V = V_{0}e^{-\lambda \psi}$, where $\lambda = 1$ have been shown. The left panel represents the evolution within the spherical overdense patch, while the right panel depicts the evolution in the background universe. The parameter $m$ is fixed at $m = 2 \times 10^{4} u$, and the Hubble parameter at initial time $t_{i}$ is $H(t_{i})=2u$, where $u=10^{-25} {\rm cm}^{-25}$.  For a fixed coupling constant $q = 0.01u^{2}$, the Red and Green curves represent dark energy potentials with $V_{0} = 0.05u^{2}$ and $V_{0} = 0.1u^{2}$, respectively. The Blue curve corresponds to $q = 5000u^{2}$ and $V_{0} = 0.1u^{2}$.}
\label{qpsis}
\end{figure*}
\subsubsection{The case where $q\psi^2 \ll m^2$}

When $q\psi^2 \ll m^2$, for a certain period, in the first approximation we can neglect the interaction term in the Klein-Gordon equation for $\phi$. In that case, the dynamical equation for $\phi$ becomes similar to the non-interacting equation of motion of $\phi$, whose solution we know in terms of the scale-factor. In such a case, the relevant dynamical equations are:
\begin{equation}\label{Friedmaneqaveragepsi1}
\frac{3\dot{a}^2}{a^2}+\frac{3k}{a^2} = \langle\rho_{\phi}\rangle + \rho_{\psi} + q\langle \phi^2\rangle\psi^2,
\end{equation}
\begin{equation}\label{Friedmaneqaveragepsi2}
-2\frac{\ddot{a}}{a} - \frac{\dot{a}^2}{a^2} - \frac{k}{a^2} = \langle P_{\phi}\rangle  +P_{\psi} - q\langle \phi^2\rangle\psi^2,
\end{equation}
and
\begin{eqnarray}\label{phiderivativeeqn}
\epsilon\ddot{\psi}+3H\epsilon\dot{\psi}-\lambda V_{0} e^{-\lambda \psi}+2q\langle\phi^{2}\rangle\psi=0.
\end{eqnarray}
Here we have only two independent unknown variables, $a(t)$ and $\psi(t)$, whereas the other variables can be expressed in terms of these two. Therefore to get the complete dynamics of the system, we need to consider only two equations out of the three. We are considering here Eqs. (\ref{Friedmaneqaveragepsi1}) and (\ref{Friedmaneqaveragepsi2}) to solve the system as they can represent two fundamental equations in the present case \cite{Bhattacharya:2024ngy}.

Using these equations we obtain one of the important equations of this paper. The second-order differential equation of $\psi$ as a function of $a$ is given as: 
\begin{widetext}
\begin{eqnarray}\label{psi(a)equation}
&&V_{0}e^{-\lambda\psi}\left(3\lambda\psi_{,a}a-4\epsilon \psi_{,a}^{2}a^{2}+\frac{\psi_{,a}^{4}a^{4}}{2}-\epsilon \psi_{,a}\psi_{,aa}a^{3}-\frac{\lambda \epsilon \psi_{,a}^{3}a^{3}}{2}\right)-\frac{5\epsilon \langle\rho_{\phi}\rangle \psi_{,a}^{2} a^{2}}{2}-\epsilon \langle\rho_{\phi}\rangle \psi_{,a} \psi_{,aa}a^{3}+\frac{\langle\rho_{\phi}\rangle \psi_{,a}^{4}a^{4}}{4}\nonumber\\&+&k(3\epsilon \psi_{,a} \psi_{,aa}a-\psi_{,a}^{4}a^{2}+9\epsilon \psi_{,a}^{2})+\frac{q\langle\rho_{\phi}\rangle\psi}{m^{2}(1+\frac{9}{8} \frac{H^{2}}{m^{2}})}\left(9\psi-\epsilon\psi\psi_{,a}^{2}a^{2}-\epsilon\psi_{,a}^{3}a^{3} -6\psi_{,a}a+\frac{\psi\psi_{,a}^{4}a^{4}}{2}-\epsilon\psi\psi_{,a}\psi_{,aa}a^{3}\right)=0\,\,.\nonumber\\
\end{eqnarray}    
\end{widetext}
Details of the derivation are provided in Appendix \ref{Appendix}. In the above equation, a comma followed by two scale-factor symbols specify a second derivative with respect to $a$.

The combined solutions of Eqs. (\ref{psi(a)equation}) and (\ref{Friedmaneqaveragepsi1}) produces the dynamics of gravitational collapse as well as the background expansion.
When we use $k=0$ in the above equations the equations become appropriate for the background. For the background, one must note that all the scale-factor expressions must be replaced by $\bar{a}$. The Eq. (\ref{psi(a)equation}) and Eq.~(\ref{Friedmaneqaveragepsi1}) have to be solved for four separate cases. By setting $\epsilon = 1$ and $\epsilon = -1$, we obtain the equations corresponding to the quintessence and phantom dark energy models, respectively. Furthermore, for any chosen dark energy model, selecting $k = 0$ or $k = 1$ will yield the equations applicable to the background universe and the over-dense region, respectively.



To obtain the dynamics of the collapsing system and the background universe we focus on a scenario where an overdense region of dark matter starts with a density contrast of $\delta_{m_i} = 10^{-3}$ relative to the background. Due to this overdensity, the region will eventually undergo a turnaround whereas the background universe will continue to expand. To solve the system, we have to choose the initial conditions judiciously.

In the starting, we will fix the parameters: $q$, $V_{0}$, $C_{\pm}$, $m$ and $\lambda$. The initial conditions of the present model depends upon the values of the parameters used in the model. All the parameters do not have the same physical dimension. In the geometrical system of units, the scalar field $\phi$ and $\psi$ and $\lambda$ are dimensionless whereas $a$ has the dimension of length, $q$, $m$ and $V_{0}$ have the dimension of inverse length squared. Here $C_{\pm}$ have the dimensions of $\text{length}^{3/2}$. We have already mentioned before that we will work with $ m = 200\times 10^{-23} \, \text{cm}^{-1} $. In our work, we will consider a unit $u=10^{-25} \, \text{cm}^{-1}$ and use it to define the parameter values. This unit is introduced so that the various variables and parameters in our theory can be expressed in terms of numbers that are not too small.

The present problem requires three initial conditions: the initial scale-factor, the initial quintessence (or phantom) field value and lastly the derivative of the quintessence (or phantom) field with respect to the scale-factor. After fixing the  parameters we set the initial scale-factor for the overdense region as $a_i = 1/u$. Since the dark matter energy density evolves as $\rho_{\phi} \propto 1/a^3$ and the background dark matter energy density also evolves as $\bar{\rho}_{\phi} \propto 1/\bar{a}^3$, as shown in Section IV, the initial dark matter  density contrast can be written as,  
$$\delta_{m_i} = \left(\frac{\bar{a}_{i}}{a_{i}}\right)^3 - 1.$$  
Given $\delta_{m_i} = 10^{-3}$, solving for $\bar{a}_{i}$ yields the initial scale-factor of the background as $\bar{a}_i = 1.0003/u$.  In this analysis we assume that the scalar field $\phi$ has the same value inside and outside of the overdense spherical perturbation, i.e. the scalar field remains continuous across the spherical boundary of the overdense region.  We also assume the same kind of continuity for the scalar field $\psi$ and consequently we do not require to fix the initial values of $\psi$ separately for the background and the overdense regions. We only require to give the initial data in the $\psi$ sector at the initial time or the initial value of the scale-factor.

The conditions for $\psi(a_{i})$ and $\psi_{, a}(a_{i})$ must be such that it produces sufficient acceleration for the background universe and causes a gravitational collapse of the overdense patch. The possible values satisfying these conditions are shown in Figs.~[\ref{phiprimevsphiplotbackground}] and [\ref{phiprimevsphiplotinside}] for the quintessence and phantom fields. In Fig.~[\ref{phiprimevsphiplotbackground}] we have plotted the background values of the field and the field derivative with respect to the scale-factor for the quintessence case (left figure) and for the phantom case (right figure). The shaded regions in these plots are made up of initial data points capable of producing accelerated expansion in the background universe.
Similarly  in Fig.~[\ref{phiprimevsphiplotinside}] we have plotted the values of the field and its derivative, in the overdense region, with respect to the scale-factor for the quintessence case (left figure) and for the phantom case (right figure). 
The shaded regions in these plots are made up of initial data points capable of producing gravitational collapse of the overdense spherical patch. As the initial conditions on the field value of $\psi$ and its derivative with respect to the scale-factor are assumed to be the same, for the background and the overdense regions, we choose the initial quintessence field data from the intersecting regions of the plots in the left panels of Figs.~[\ref{phiprimevsphiplotbackground}] and [\ref{phiprimevsphiplotinside}]. For the phantom field a similar procedure is applied for the plots on the right panels of Figs.~[\ref{phiprimevsphiplotbackground}] and [\ref{phiprimevsphiplotinside}].

If the condition of $\psi(a_{i})$ and $\psi_{, a}(a_{i})$ for the over-dense region and $\psi(\bar{a_{i}})$ and $\psi_{,\bar{a}}(\bar{a_{i}})$ for the background are chosen from appropriate intersections of the allowed regions of both Figs.~[\ref{phiprimevsphiplotbackground}, \ref{phiprimevsphiplotinside}] we will have an overdense spherical region proceeding towards gravitational collapse in the background of an accelerated phase of expansion of the universe.

\begin{figure*}
\centering
\begin{minipage}[t]{.4\textwidth}
\centering
\includegraphics[width=70mm,height=52mm]{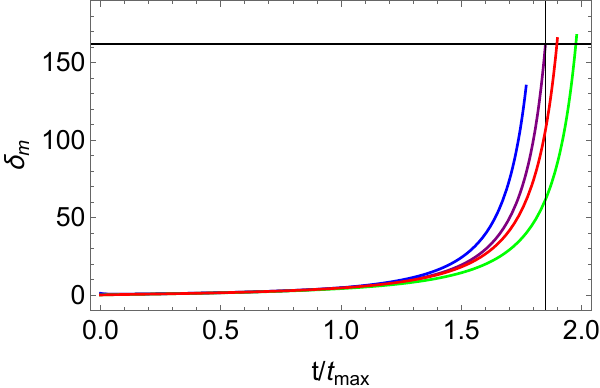}
\caption{The figure illustrates the evolution of the density contrast of dark matter up to the point of virialization (till $a(t)=a_{\rm max}/2$). Here dark matter is non-minimally coupled with quintessence-like dark energy. The color coding is as follows: purple curve represents the top-hat collapse scenario without any dark energy. For a fixed coupling constant $q=0.01u^{2}$, red curve corresponds to $V_{0}=0.05u^{2}$, green curve corresponds to $V_{0}=0.1u^{2}$. The blue curve corresponds to a different coupling constant $q=5000u^{2}$ and $V_{0}=0.1u^{2}$, where $u=10^{-25} \text{cm}^{-1}$.}
\label{densitycontrastplotq}
\end{minipage}\qquad
\begin{minipage}[t]{.4\textwidth}
\centering
\includegraphics[width=70mm,height=52mm]{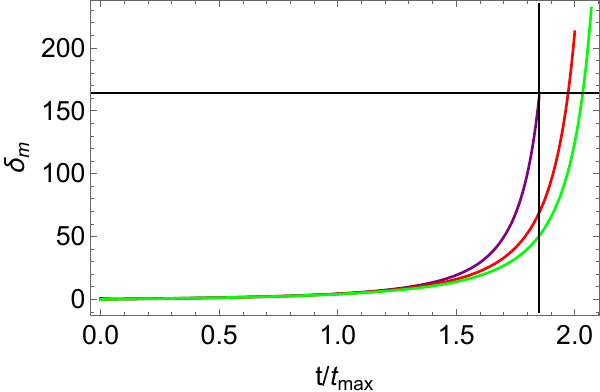}
\caption{The figure illustrates the evolution of the density contrast of dark matter up to the point of virialization (till $a(t)=a_{\rm max}/2$). Here dark matter is non-minimally coupled with phantom-like dark energy. The color coding is as follows: purple curve represents the top-hat collapse scenario without any dark energy. For a fixed coupling constant $q=0.01u^{2}$, the red curve corresponds to $V_{0}=0.08u^{2}$, green curve corresponds to $V_{0}=0.1u^{2}$, where $u=10^{-25} \text{cm}^{-1}$.}
\label{densitycontrastplotp}
\end{minipage}
\end{figure*}
\subsubsection{The cases where $q\psi^2 \gg m^2$ and $q\psi^2 \sim m^2$}
For the cases where $q\psi^2 \gg m^2$ or $q\psi^2 \sim m^2$, we have $\left( m^2 + 2q\psi^2 \right)^{\frac{1}{2}} \gg H$, as we are only considering situations where $m \gg H$. In this regime, the terms involving $q\psi^2$ cannot be neglected, and they modify Eq.~(\ref{m2term}) as follows:
\begin{equation}
    \dot{\alpha}^2 - m^2 - 2q\psi^2(t) = 0,
\end{equation}
which implies that $\alpha(t)$ may not depend linearly on $t$. Instead, its functional form is entirely determined by $\psi(t)$. Consequently, in these cases, $\sin(\alpha(t))$ and $\cos(\alpha(t))$ may not oscillate with time, and therefore, $\phi(t)$ may also cease to be a rapidly oscillating function on the cosmological timescale $\frac{1}{H}$. As a result, the collapsing system lacks an initial dust component, meaning we cannot claim our results as a direct generalization of the standard top-hat collapse model. The higher the initial value of $q\psi^2$, the further the system deviates from a dust-like dark matter scenario. Since we explicitly aimed to model dark matter as dust, we have avoided calculations in this regime.

In the standard treatment of oscillating scalar field dark matter (in the absence of interaction with dark energy) it is assumed that the cosmological Hubble parameter diminishes in time and at some time in the evolution of the universe $m\gg H$ and we have dust like behaviour in the dark matter sector \cite{Turner:1983he}. Our work shows that if the dark matter constituent interacts with the quintessence (or phantom) field, such an assumption may not work. Whether the dark matter sector behaves as dust or as a fluid with nonzero pressure depends upon the value of $\psi(t)$ and the coupling constant $q$.  If 
$q\psi^2 \gg m^2$ then the dark matter sector will in general have nonzero pressure. 

In the present work we have assumed $q\psi^2 \ll m^2$ and worked with the traditional dust like dark matter sector. We have checked that once initially we assume $q\psi^2 \ll m^2$, the condition remains valid throughout the time period of our concern showing internal consistency of the system. This fact is highlighted in Fig.~[\ref{qpsis}] in the background as as well as inside the collapsing spacetime.
On the other hand, if $q\psi^2 \sim m^2$ initially, it is seen that the condition does not remain valid after a certain time period showing that such a condition contradicts the dust-like behavior of the dark sector. Consequently, we have not shown any result in the limit where $q\psi^2 \sim m^2$.  
\section{Non-minimal coupling and gravitational collapse: Results}
\label{Density Contrast}

Here we have started with a slightly over-dense, spherical region with a slightly different matter density compared to the background. This spherical region detaches from the background expansion and expands at a different rate from the background due to the density contrast. After some time this over-dense region will turn around leading to a collapse and the collapse continues till the time when the system stabilizes into a virialized state. The exact process in which the virialization occurs cannot be predicted from a general relativistic perspective. It is assumed that the end equilibrium state is a spherically symmetric, static spacetime. In \cite{Joshi1, Dey2, Dey3, Koushiki1}, the authors define a general relativistic equilibrium state that can be attained as the end state of gravitational collapse for general spherically symmetric compact objects. However, they do not address whether this general relativistic equilibrium state is equivalent to the Newtonian virialized state, leaving this question open for further investigation.

Let's say the initial density of the over-dense region, 
\begin{eqnarray}
\rho_{\phi_{i}} \equiv \bar{\rho}_{\phi_{i}}(1+\delta_{m_{i}}),
\end{eqnarray}
where $\rho_{\phi_{i}}$ and $\bar{\rho}_{\phi_{i}}$ are the initial dark matter energy density for the inside, overdense region, and the background, respectively and $\delta_{m_{i}}$ is the initial density contrast. In the standard top-hat collapse the collapsing spherical region has only dust inside it. In top-hat collapse, the density contrast starts with a minimal value (around zero) and reaches a value of approximately $170-180$ at the time of virialization. In our work, the scalar field $\phi$ has been used to represent dark matter candidate and we have shown that if $\phi$ is only present in the system then the collapse will behave like the ideal top-hat collapse. In such a case we obtain the above-mentioned density contrast at the time of virialization. We will specify more about density contrast in the dark matter sector, when $\phi$ is coupled to $\psi$, at the end of this section.

The fixed parameter values used in this paper are 
$\lambda=1$, $m= 2 \times 10^{4} u$, and $C_{\pm}=2\times 10^{-4}/ u^{3/2}$.  These values are chosen in such a way that in conjunction with the initial conditions we get reasonable results showing gravitational collapse of an isolated spherical overdense patch when the background universe goes through an accelerated expansion phase.  $V_0$ and $q$ differ in some of the graphs presented, so we do not specify their value here, the particular values of $q$ and $V_0$ used to generate the results can be seen from the captions of the various figures. These values directly control the effect of dark energy on the collapse process. The possible values of $\psi(a_{i})$ and $\psi_{, a}(a_{i})$ or $\psi(\bar{a_{i}})$ and $\psi_{,\bar{a}}(\bar{a_{i}})$ satisfying these conditions are shown in Figs.~[\ref{phiprimevsphiplotbackground}, \ref{phiprimevsphiplotinside}]. We choose the same initial values for inside and the background, therefore the values of $\psi(a_{i})$ and $\psi_{, a}(a_{i})$ or $\psi(\bar{a_{i}})$ and $\psi_{,\bar{a}}(\bar{a_{i}})$ are obtained from the intersection of the allowed regions. Regions outside the intersection region will not produce a collapse or if it does the background spacetime will not have appropriate accelerating expansion. For quintessence field $\psi_{, a}(a_{i})$ and $\psi_{,\bar{a}}(\bar{a_{i}})$ are kept fixed at $0.001u$ for all the figures in Fig.~[\ref{Time evolution for quintessence field}]. When $q=5000u^{2}$ and $V_{0}=0.1u^{2}$ then $\psi(a_{i})=\psi(\bar{a_{i}})=6$. Again fixing $q=0.01u^{2}$ we have plotted two figures corresponding to $V_{0}=0.1u^{2}$ and $V_{0}=0.05u^{2}$, in both the cases $\psi(a_{i})=\psi(\bar{a_{i}})=3.5$. For phantom field the results are shown in Fig.~[\ref{Time evolution for phantom field}].  In this case $q=0.01u^{2}$, $\psi(a_{i})=\psi(\bar{a_{i}})=3.2$ and $\psi_{, a}(a_{i})=\psi_{,\bar{a}}(\bar{a_{i}})=0.005u$ are fixed for all the figures while only the potential form varies for the various figures as $V_{0}=0.08u^{2},0.1u^{2}$. When the dark energy constituent is made up of a phantom field, it is seen that the situation becomes extremely sensitive to the magnitude of the coupling constant $q$. Increasing the magnitude of $q$ slightly can lead to large energy exchange between the phantom sector and the dark matter sector making the phantom energy density negative. As we demand that all components of energy density remain positive we do not have the freedom of changing $q$ appreciably in this sector. As a consequence the plots in  Fig.~[\ref{Time evolution for phantom field}] are plotted for a single value of $q$.

The dynamics of the collapse of the spherical overdense patch is shown in the various plots in Fig.~[\ref{Time evolution for quintessence field}] when the dark energy component is the quintessence field. In Fig.~[\ref{Time evolution for quintessence field}(b)] we have plotted the collapsing scale-factor with respect to time for various values of $V_0$. The purple curve specifies the result of pure top-hat collapse (in the absence of any dark energy component). To see the effect of dark energy we have drawn a vertical line at a time where the top-hat collapse virialize and another horizontal line specifying the ratio $a/a_{\rm max}$ at virialization for the top-hat collapse model. The intersection of these lines with the curves qualitatively show that the presence of dark energy can alter the results of top-hat like collapse. In the present case it is very difficult to find out the exact time when the combined system virializes. We expect the combined system to virialize near the top-hat virialization time but an exact value for the virialization time cannot be calculated analytically. The results indicate that the quintessence field can produce both unclustered or clustered dark energy, the result depends upon the magnitude of $q$.  For a relatively small value of $q$ the plots in $\rho_\psi/\bar{\rho}_{\psi}$ and $\omega_\psi$ show  that the quintessence field dynamics, in the background and inside the overdense patch, approximately behaves similarly. This result shows approximate unclustered dark energy.  In this case one can see that the results depend sensitively on the magnitude of $V_0$. The effective equation of state $\omega_t$ shows that the effective fluid inside the overdense patch behaves approximately as dust and consequently exerts practically feeble pressure on the boundary region. The background spacetime shows accelerating expansion as demanded from the system.  The plots show that the results do depend upon the choice of $V_0$ and altering $V_0$ can have very different outcomes. If the value of $q$ is increased then the results are depicted in the blue colored plots. The dynamical behaviour of the system changes drastically if $q$ is increased by a fair amount. In this case it is seen that the virialized state is reached before the standard top-hat virialization time. Moreover the  plots in $\rho_\psi/\bar{\rho}_{\psi}$ and $\omega_\psi$ show  the signatures of clustered dark energy. As a result we can say that the variation of $q$ can produce qualitative changes in the results of gravitational collapse where the dark sector is interacting non-minimally. Of course, all the results shown are true for some chosen set of parameter values and the results may differ when these parameters change. Our work gives a very rich set of possibilities.

The collapse dynamics in the presence of the phantom field are specified in the various plots in Fig.~[\ref{Time evolution for phantom field}]. The results are interpreted similarly. In this case as because the phantom energy density changes drastically with a small change in the magnitude of $q$ and it is difficult to obtain a proper collapse process with $\rho_\psi \ge 0$,  we have plotted all the curves for the same value of $q$ for which all the energy densities are positive definite. The various other curves of interest are obtained by varying the magnitude of $V_0$. The purple curve in Fig.~[\ref{Time evolution for phantom field}(b)] represents the scale-factor dynamics in the top-hat like collapses as before. The other curves and their intersections with the vertical and horizontal straight lines in Fig.~[\ref{Time evolution for phantom field}(b)] directly show that non-minimal coupling in the dark sector affects the gravitational collapse. The plots show that in this case the dark energy is approximately unclustered and the effective equation of state of the combined fluid inside the overdense patch is near zero showing dust-like behavior. The effective equation of state in the background is smaller than $-1$ as expected in a universe where the dark energy component is phantom-like.    

When we include dark energy and the interaction part in the system, the evolution of density contrast also changes and its value depends on the value of $V_{0}$, where $V_{0}$ is the parameter in the $\psi$ field potential $V(\psi)$. In Figs.~[\ref{densitycontrastplotq}, \ref{densitycontrastplotp}] we have shown the density contrast evolution for dark matter, where quintessence and phantom-like dark energy are present in the system respectively. In both figures it shows that when the value of $V_{0}$ increases, for both quintessence and phantom fields, the density contrast also increases near the time of virialization, assuming that virialization happens in a time period which is near to the time period of virialization in the pure top-hat collapse. Consequently incorporating dark energy into the system changes the dark matter density contrast evolution inside the system, a result which shows the effect of the non-minimal coupling in the dark sector. In both the figures the purple curves represents the density contrast in a pure top-hat collapse and the horizontal and vertical lines help us to understand the difference between the three curves for a particular value of the density contrast or at a particular time.

Our results show that for some specific initial conditions and for some specific choice of parameter values the system of non-minimally coupled scalar fields can indeed produce a top-hat like collapse which in general differ from the standard top-hat collapse which takes place in the presence of dust.

\section{Discussion and conclusion}

In the present paper we have tried to model top-hat like collapse when the dark sector is composed of a light, oscillating scalar field acting as dark matter candidate and another scalar field, which can be the standard quintessence field or the phantom field, which constitutes the dark energy component. In the presence of a non-minimal interaction between the scalar fields we see that top-hat like gravitational collapse of dark matter gets affected. Our calculation fully employs the general relativistic formalism till the time of virialization. Virialization is an adhoc concept which specifies the process in which the gravitational collapse comes to a halt and after which stable structures are formed. The properties of virialization for the non-interacting dark sector is known to a certain extent but we still do not have any theory of virialization where the dark sector components interact non-minimally. In this paper we have assumed that the interacting system virializes at a time scale which is not too different from the analogous time scale for a pure top-hat collapse which happens in the absence of any dark energy candidate. 

Inclusion of non-minimal coupling in the dark sector gives rise to nonzero pressure inside the overdense spherical region undergoing gravitational collapse. To account for this nonzero pressure one has to apply proper junction conditions at the boundary of the internal spherical region. The immediate neighbourhood of the collapsing patch is modelled to be a generalized Vaidya spacetime. As soon as the overdense region, with a minute density contrast with respect to the background, detaches from the background expansion this Vaidya spacetime also comes into existence. Our results show that the internal spacetime undergoing the gravitational collapse may have very small pressure (for the particular initial conditions and parameter values chosen) and in such a case the external spacetime may act more like a Schwarzschild spacetime. 

In this paper we have modelled the dark matter sector as an oscillating field, oscillating rapidly in the cosmological time scale. To have such an oscillation phenomena one requires the mass parameter of the field to be very small, in our case this mass parameter is of the order of $10^{-26}$eV. By properly adjusting the other parameters in the theory one may increase the value of this parameter but the increase will not be appreciably large. Our model predicts an ultra low mass dark energy candidate. The rapidly oscillating part of the dark matter field is appropriately averaged out and consequently it brings in new features in the theory. An upper bound on $q\psi^2$ is essential for the averaging process in the dark matter sector to work. 

The theory on which the present work is based is heavily constrained. The various requirements, which constrain the system, are specified as:
\begin{enumerate}
\item The requirement of an accelerating expansion in the background FLRW spacetime.

\item The requirement that a spherical patch experiences a gravitational collapse in the aforementioned background.

\item The requirement that the phantom energy density $\rho_\psi$ $(\epsilon=-1)$ remains positive in the background as well as in the spherical overdense region during the whole process. One must note that $\langle \rho_\phi \rangle$ is always positive and $\rho_{\rm int}$ is also positive for positive $q$.

\item The requirement that $m\gg H$ and $m \gg \bar{H}$ initially.

\item The requirement that $q\psi^2\ll m^2$ throughout the collapsing time period.
\end{enumerate}
All the above listed requirements have to be satisfied and consequently it is very difficult to obtain the desired collapsing phase in a background universe undergoing accelerated expansion in the presence of non-minimally interacting scalar fields in the dark sector. To address all these requirements the model we have presented requires a six dimensional parameter space. Non-minimal interactions in general constrains the system and allows top-hat like collapse for a certain range of values of $\psi$, the range is given by $\psi\ll\psi_c $ where $\psi_c^2=m^2/q$. If $\psi \ge \psi_c$ then top-hat like collapse will not be viable or will not remain stable. In such cases one has to find some alternative process in which nonlinear perturbations develop in the late cosmological era. 

We have shown that in presence of non-minimal interaction between the scalar fields, in the dark sector, one can indeed use general relativistic techniques to model top-hat like collapse up to a phase of virialization. Although in our work the dimensional coupling constant, between the dark sector components, is assumed to be small, the effect of the non-minimal coupling in the dark sector affects the gravitational collapse process and qualitatively we get generalized top-hat collapse like phenomena. The results are sensitive to the choice  of the parameters. In the paper we have showed the dependence of the results on the choice of values of $V_0$. Due to a large number of parameters our work can have various kinds of results depending on the parameters chosen. We have only worked with some values in this wide parameter space and we expect that future workers can find more interesting results from our framework when they work in some other regions of the parameter space. In the parameter space we have chosen, we have obtained an overdense spherical collapsing region where dark energy is approximately unclustered and the spherical region has a feeble effective pressure. Lastly, we have proposed a new way to deal with interactions between scalar fields, in the cosmological sector, where one of the scalar degrees of freedom has time averaged behaviour while the other scalar field dynamics does not show any oscillating behaviour in the cosmological time scale.

\appendix

\section{}\label{Appendix}
Using Eq. (\ref{Friedmaneqaveragepsi1}), we get
\begin{eqnarray}\label{five}
\dot{a}=\pm\sqrt[]{\frac{(\langle\rho_{\phi}\rangle+\rho_{\psi}+\rho_{\text{int}})a^{2}}{3}-k}\, .
\end{eqnarray}
Differentiating Eq. (\ref{five}) with respect to coordinate time (t) we get
\begin{eqnarray}\label{six}
\ddot{a}=\frac{a}{3}\left[\langle\rho_{\phi}\rangle+\rho_{\psi}+\rho_{\text{int}}+\frac{a}{2}(\langle\rho_{\phi}\rangle_{, a}+\rho_{\psi_{,a}}+\rho_{\text{int}_{,a}})\right]\, ,\nonumber\\
\end{eqnarray}
where $\langle\rho_{\phi}\rangle_{, a}$, $\rho_{\psi,_{a}}$, and $\rho_{\text{int}_{, a}}$ are derivatives of the dark matter energy density, dark energy density and the interaction part energy density respectively, with respect to the scale factor $a$.
Now from Eq. (\ref{rhoforpsi}), we get,
\begin{eqnarray}\label{psi,aequation}
\langle\rho_{\phi}\rangle_{, a}=-\frac{3\langle\rho_{\phi}\rangle}{a}.
\end{eqnarray}
In this case we can write
\begin{eqnarray}\label{seven}
\rho_{\psi}+P_{\psi}=\epsilon\psi_{,a}^{2}\dot{a}^2 &~\text{and}~& P_{\psi}=\rho_{\psi}-2V(\psi)\, .
\end{eqnarray}
Substituting the expression of $\dot{a}$ from Eq.~(\ref{five}) in the Eq.~(\ref{seven}) we get
\begin{equation}\label{nine}
\rho_{\psi}\left(1-\frac{\epsilon\psi_{,a}^{2}a^{2}}{3}\right)-\left(\langle\rho_{\phi}\rangle + \rho_{\rm int}\right)\frac{\epsilon\psi_{,a}^{2}a^{2}}{3}+P_{\psi}+k\epsilon\psi_{,a}^{2}=0\, .
\end{equation}
Using the value of $P_{\psi}$ from Eq.~(\ref{seven}) in Eq.~(\ref{nine}) we get
\begin{equation}\label{ten}
\rho_{\psi}=\frac{\frac{(\langle\rho_{\phi}\rangle+\rho_{\rm int})\epsilon\psi_{,a}^{2}a^{2}}{6}+V(\psi)-\frac{k\epsilon\psi_{,a}^{2}}{2}}{\left(1-\frac{\epsilon\psi_{,a}^{2}a^{2}}{6}\right)}\, .
\end{equation}
Using Eqs.~(\ref{Friedmaneqaveragepsi2}), (\ref{five}), (\ref{six}), (\ref{seven}) we get
\begin{eqnarray}\label{tweleve}
\rho_{\psi,a}&=&-\left[\left(\frac{\langle\rho_{\phi}\rangle}{a}+\frac{\rho_{\rm int}}{a}\right)\left(3+\epsilon\psi_{,a}^{2}a^{2}\right)+\rho_{\psi}\epsilon\psi_{,a}^{2}a+\frac{3P_{\rm int}}{a}\right.\nonumber\\
&&\left.-\frac{3k\epsilon\psi_{,a}^2}{a}\right]
-\langle\rho_{\phi}\rangle_{, a}-\rho_{\text{int},a}\, .
\end{eqnarray}
Now we differentiate Eq.~(\ref{ten}) with respect to $a$ and use Eq.~(\ref{tweleve}) to obtain:
\begin{eqnarray}  \label{eleven}
&-&\left[\frac{(\langle\rho_{\phi}\rangle+\rho_{\text{int}})(3+\epsilon\psi_{,a}^{2}a^{2})+\rho_{\psi}\epsilon\psi_{,a}^{2}a^{2}+3P_{\text{int}}-3k\epsilon\psi_{,a}^{2}}{a}\right]\nonumber\\&-&\langle\rho_{\phi}\rangle_{, a}-\rho_{\text{int},a}=\frac{1}{3(1-\frac{\epsilon\psi_{,a}^{2}a^{2}}{6})^{2}}\{3V_{,\psi}\psi_{,a}-3k\epsilon\psi_{,a}\psi_{,aa}\nonumber\\&+&\frac{(\langle\rho_{\phi}\rangle_{, a}+\rho_{\text{int},a})\epsilon\psi_{,a}^{2}a^{2}}{2}-\frac{(\langle\rho_{\phi}\rangle_{, a}+\rho_{\text{int},a})\epsilon^{2}\psi_{,a}^{4}a^{4}}{12}\nonumber\\&+&(\langle\rho_{\phi}\rangle+\rho_{\text{int}})\epsilon\psi_{,a}\psi_{,aa}a^{2}-\frac{\epsilon V_{,\psi}\psi_{,a}^{3}a^{2}}{2}+\epsilon V(\psi)a\psi_{,a}^{2}\nonumber\\&+&\epsilon V(\psi)a^{2}\psi_{,a}\psi_{,aa}+(\langle\rho_{\phi}\rangle+\rho_{\text{int}})\epsilon \psi_{,a}^{2}a-\frac{k\epsilon^{2}\psi_{,a}^{4}a}{2}\}\,\,.\nonumber\\
\end{eqnarray}

Now using Eqs. (\ref{phipm}), (\ref{phisqeq}), (\ref{potentialpsi}), (\ref{rhoint}), (\ref{pint}) and (\ref{psi,aequation}), in Eq. (\ref{eleven}), we can derive the second-order differential equation of $\psi$ as a function of a.


\end{document}